\documentstyle[epsf,11pt,mypp]{article}

\newcommand{\comment}[1]{}

\def\be{\begin{equation}}
\def\bel#1{\begin{equation}\label{eq:#1}}
\def\ee{\end{equation}}
\def\bea{\begin{eqnarray}}
\def\beal#1{\begin{eqnarray}\label{eq:#1}}
\def\eea{\end{eqnarray}}
\def\eqref#1{\ref{eq:#1}}

\def\eqn#1{equation~(\ref{eq:#1})}

\def\eqnp#1{eq.~[\ref{eq:#1}]}

\def\fig#1{Figure~\ref{fig:#1}}

\def\lci{LC1}
\def\lcii{LC2}

\def\etal{~\mbox{et al.}}

\def\deg{$^\circ$}
\def\kms{km/s}

\def\myvec#1{\mbox{\boldmath $#1$}}
\def\ewe{{U}}
\def\vee{{V}}
\def\wei{{W}}
\def\weiel{{w}}
\def\espec{\myvec{v}}
\def\svec{\myvec{s}}
\def\smat{S}
\def\specel{s}
\def\Cov{{C}}

\def\Mbri{{M_{\rm bri}}}
\def\Mdim{{M_{\rm dim}}}
\def\Mstar{M_{\ast}}

\def\gae{\lower 2pt \hbox{$\, \buildrel {\scriptstyle >}\over {\scriptstyle
\sim}\,$}}
\def\lae{\lower 2pt \hbox{$\, \buildrel {\scriptstyle <}\over {\scriptstyle
\sim}\,$}}

\begin{document}

\markboth{Bromley et al.}{Spectral Classification of the LCRS}
\pagestyle{myheadings}
\thispagestyle{empty}

\title{Spectral Classification and Luminosity Function of Galaxies in the 
       Las Campanas Redshift Survey}

\author{
Benjamin C. Bromley and William H. Press, 
}
\affil{
Physics Department, Harvard University, and
Harvard-Smithsonian Center for Astrophysics
}
\authoraddr{
60 Garden Street, Cambridge, MA 02138
}
\author{
Huan Lin
}
\affil{
Department of Astronomy, University of Toronto
}
\authoraddr{
60 St. George Street, Toronto, ONT M5S 3H8, Canada
}
\and
\author{
Robert P. Kirshner
}
\affil{
Harvard-Smithsonian Center for Astrophysics
}
\authoraddr{
60 Garden Street, Cambridge, MA 02138
}

\begin{abstract}

We construct a spectral classification scheme for the galaxies of the
Las Campanas Redshift Survey (LCRS) based on a principal component
analysis of the measured galaxy spectra. We interpret the physical
significance of our six spectral types and conclude that they are
sensitive to morphological type and the amount of active star
formation.  In this first analysis of the LCRS to include spectral
classification, we estimate the general luminosity function, expressed
as a weighted sum of the type-specific luminosity functions.  In the
$R$-band magnitude range of $-23 < M \leq -16.5$, this function
exhibits a broad shoulder centered on $M \approx -20$, and an
increasing faint-end slope which formally converges on $\alpha \approx
-1.8$ in the faint limit.  The Schechter parameterization does not
provide a good representation in this case, a fact which may partly
explain the reported discrepancy between the luminosity functions of
the LCRS and other redshift catalogs such as the Century Survey
(Geller\etal~1997). The discrepancy may also arise from environmental
effects such as the density-morphology relationship for which we see
strong evidence in the LCRS galaxies.  However, the Schechter
parameterization is more effective for the luminosity functions of the
individual spectral types.  The data show a significant, progressive
steepening of the faint-end slope, from $\alpha \approx +0.5$ for
early-type objects, to $\alpha \approx -1.8$ for the extreme late-type
galaxies. The extreme late-type population has a sufficiently high
space density that its contribution to the general luminosity function
is expected to dominate fainter than $M = -16$.  We conclude that an
evaluation of type-dependence is essential to any assessment of the
general luminosity function.

\end{abstract}


\section{Introduction}\label{sect:intro}

Classification of galaxies has long been regarded as an important step
toward understanding galaxy formation and evolution.  There is
considerable evidence that the two main methods of
classification---morphological and spectral---both directly reflect
physical properties of galaxies (Roberts \&
Haynes~\cite{RobHay94}). Indeed it was clear early on that these two
methods yield galaxy types which are strongly correlated (Morgan \&
Mayall~\cite{MorMay57}), although recent work shows that the
correlation is not perfect (Connolly\etal~\cite{ConEtal95}; Zaritsky,
Zabludoff \& Willick~\cite{ZarZabWil95}).  In preparation for large
galaxy catalogs such as the Sloan Digital Sky Survey efforts have
begun to automate both approaches to classification.
Morphologically-based schemes must of necessity deal with pixel
images, and the most promising technique employs artificial neural
networks (e.g., Odewahn\etal~\cite{OdeEtal96}; Folkes, Lahav \&
Maddox~\cite{FolLahMad96}). Spectral schemes, on the other hand,
require no more information than is already available in a typical
redshift catalog. Furthermore, the interpretation of a spectral type
is facilitated by the direct connection between spectral features and
the physics of the stars and gas within the galaxies.

The information contained in a set of galaxy spectra can be extracted
in an efficient manner by a principal component method (e.g.,
Connolly\etal~\cite{ConEtal95}). The idea is to identify features of
the spectra which show the greatest variation from galaxy to galaxy,
so that a classification scheme can reflect those physical properties
which maximally distinguish galaxies.  We here apply the principal
component method to analyze the galaxies in the Las Campanas Redshift
Survey, a large multiple-strip survey with over 25,000 measured
spectra.

The goal of the work described here is not only to define spectral
types but to begin discerning general physical properties of the
galaxies in each type. We focus on the luminosity function, an
important constraint for models of cosmic structure formation.  We
begin with a review of the LCRS catalog (\S2), then, in Section~3, we
outline the principal component method and compare our classification
scheme with previous work.  In Section~4 we present our analysis of
the luminosity function according to spectral type.

\section{The Redshift Data}\label{sect:data}

The Las Campanas Redshift Survey is described in detail in
Shectman\etal~(\cite{SheEtal96}; hereafter \lci). Lin et al.\
(\cite{LinEtal96}; hereafter \lcii) provide additional details related
to galaxy selection.  We summarize characteristics of the survey
briefly in this section.

The survey consists of $\sim$26,000 galaxies in six sky strips, three
strips in the northern Galactic cap region and three in the southern
region.  Each strip runs approximately 80\deg\ across the sky in right
ascension and has a width in declination of $\sim$1.5\deg. The mean
strip declinations are --3\deg, --6\deg\ and --12\deg\ in the northern
sample and --39\deg, --42\deg\ and --45\deg\ in the south.

Each sky strip was subdivided into 50 or so fields of square or nearly
square dimensions.  Galaxies in each field were selected on the basis
of Kron-Cousins $R$-band magnitudes; a subset of these galaxies were
chosen randomly for spectroscopic study using multiobject fiber
spectrometers of either 50 or 112 fibers.  Apparent magnitude limits
vary from field to field, with typical isophotal limits of $16.0 \leq
m < 17.3$ and $15.0 \leq m < 17.7$ for the 50-fiber and 112-fiber
fields, respectively.  Additional limits were imposed on the basis of
``central surface brightness'' of the galaxies, corresponding
approximately to the flux entering a fixed fiber aperture of
3.5$^{\prime\prime}$; the limiting central magnitude is in the range
of $m_c = 18$ to 19, depending on the isophotal magnitude.  This
additional cut (a feature common to all magnitude-limited surveys,
whether quantified or not) amounts to a 20\% reduction in the fraction
of galaxies in the 50-fiber fields and less than a 10\% reduction of
the 112-fiber fields.
%
%
We emphasize that each field has assigned to it a unique set of
parameters which include isophotal magnitude limits, central magnitude
cut, and ``sample fraction'' of the total galaxies in that field for
which spectra were observed.  As in \lcii, we take full account of
these field-to-field differences.

The individual spectra range from 3350~\AA\ to 6750~\AA\ for the
112-fiber data and from 3350~\AA\ to 6400~\AA\ for the 50-fiber
data. The wavelength resolution is $\sim 7$~\AA\ (FWHM) and the
sampling rate is 2.5~\AA\ per channel. As discussed below, we consider a
narrower range of wavelengths to allow the redshifted spectra to be
rescaled into a common waveband in the rest frame.

We work with the full catalog of 25,327 galaxies to perform the
spectral classification in the next section. Following \lcii, our
analysis of the luminosity function in \S\ref{sect:lf} makes use of
only those galaxies with absolute magnitudes between --23 and --16.5
and with redshifts such that $1,000 \mbox{\ \kms} \leq c z <
60,000$~\kms.  We also exclude the 50-fiber data because of their more
stringent surface brightness cuts.  This leaves 18,105 objects, 8228
in the north and 9877 in the south.
%
%
The same labeling scheme appears here as in \lcii, so, for example, N112
refers to the northern 112-fiber data.

\section{Spectral Classification}\label{sect:spec}

Toward quantitative classification of the LCRS catalog we seek a
representation of the spectra which accentuates the differences
between individual galaxies.  Such a representation is afforded by
singular-value decomposition (SVD) as part of a principal component
analysis (see Kendall~\cite{Ken75}).  The overall strategy is to
express the spectra in terms of a series of templates---following
Connolly\etal~(\cite{ConEtal95}) we call them ``eigenspectra''---and
to select those which are most sensitive to galaxy-to-galaxy
variations. The coefficient obtained by projecting a galaxy's spectrum
onto such an eigenspectrum can serve as its spectral
type. Singular-value decomposition gives the general prescription for
how to build the eigenspectra; principal component analysis
demonstrates that they indeed characterize those properties of the
galaxies which optimally distinguish them.

\subsection{Singular-value decomposition}\label{ssect:svd}

Let the vector $\svec_i$ represent the spectrum of the $i^{\rm th}$
galaxy in a catalog of $N$ objects.  The components of the vector,
$\specel_{ij}$, are integrated fluxes in frequency channels, where $j$
is the channel index.  The full catalog then is a matrix $\smat$ of
size $N \times M$ where $N$ is 25,327 and the number of channels, $M$,
is 800.  We first preprocess the spectra in this matrix by weighting
each wavelength channel using a smoothed version of the mean spectrum
in the observatory-frame (this step downweights those wavelengths
where the instrumental response is low).  We then normalize the
weighted spectra to some constant flux value.  Next, the data are
shifted into the rest frame and high-pass filtered to remove
fluctuations in the spectra which have power on scales of
$\sim$1000~\AA\ or more. Finally we subtract off the average of the
filtered spectra---we want the power in the eigenspectra to reflect
differences between galaxies, not their similarities.  It is this
normalized, shifted, high-pass filtered entity which we call the
catalog, $\smat$.

We can write the matrix $\smat$ in the form 
\bel{svd}
       \smat = \ewe \wei \vee^T \ ,
\ee
where $\ewe$ is a row-orthogonal $M \times N$ matrix, and $\wei$ and
$\vee$ are both $M \times M$; the former is diagonal with elements
$\weiel_i \geq 0$ and the latter is orthonormal
(cf.~Press\etal~\cite{PreEtal92}, \S 2.6).  This equation is the
singular-value decomposition of $\smat$. It is generally valid for all
matrices provided that the number of columns does not exceed the
number of rows, i.e., $N \geq M$.

Notice that any single spectrum $\svec_i$ from the catalog can be
easily reconstructed from the matrices $\ewe$, $\wei$ and $\vee$ in
\eqn{svd}:
\bel{reconstruct}
    \svec_i = \sum_{j = 1}^{M}  \gamma_{ij} \espec_j
\ee
where $\gamma_{ij}$ is the $ij^{\rm th}$  element of the product $\ewe 
\wei$,
and the vector $\espec_j$ is the $j^{\rm th}$ column of $\vee$.  This
property highlights the special significance of $\vee$: its columns
define orthonormal vectors $\espec_1$...$\espec_M$, the eigenspectra
which form the basis used in a principal component analysis. The
defining characteristic of such a basis is that it diagonalizes the
covariance matrix of the catalog,
\bel{cov}
    \begin{array}{rcl}
    \Cov & \equiv & \smat^T\smat \ , \\
         & = & \vee \wei^T \ewe^T \ewe \wei \vee^T \ , \\
         & = & \vee \wei^2 \vee^T  \ ,
    \end{array}   
\ee
(note that $\ewe^T = \ewe^{-1}$). Projecting the expressions in
\eqn{cov} onto the eigenspectra explicitly demonstrates that $\Cov$ is
diagonalized, leaving the eigenvalue equation
\bel{espec}
     \Cov \espec_i = \weiel_{i}^2 \espec_i \ ,
\ee
where $\weiel_{i}$ is the $i^{\rm th}$ diagonal element of
$\wei$ in \eqn{svd}. 

There may also be value in weighting individual data points
differently, depending on measurement errors or prior assumptions
about the data.  Thus, before calculating the SVD, one may wish to
write the elements of the catalog matrix as
\bel{pcawei}
\specel_{ij} \rightarrow g_{i} h_{j} \specel_{ij} \ ,
\ee
where $g_{i}$ and $h_{j}$ are weight factors.  A transformation of
this type can describe the effects of the ``renormalization'' which we
performed on the spectra in the LCRS catalog.

A geometric interpretation of the eigenspectra is that they are
aligned with the principle axes of the error ellipsoid corresponding
to the scatter of points in the catalog. The eigenspectra with the
largest eigenvalues (the $\weiel_i^2$ in \eqnp{espec}) indicate
directions in the space of frequency channels where there is greatest
variation between objects in the catalog.  Conversely, small
eigenvalues correspond to directions along which the data show little
variation.

The idealization that data scatter is an error ellipsoid hints that a
principal component analysis is best suited to systems where the data
correspond to $M$ statistically independent Gaussian variates.
Knowledge of the covariance matrix allows the original orthogonal
basis in which the data are collected (i.e., frequency channels) to be
rotated into a new basis (the eigenspectra) to represent the data in
terms of statistically independent coefficients (the $\gamma_{ij}$ in
\eqnp{reconstruct}).  This rotation is the Karhunen-Lo\`{e}ve
transform. In general, where higher-order correlations are possible,
these coefficients will not necessarily be statistically independent.

A data compression algorithm is evident from the geometrical
interpretation above: Simply keep only the first $m$ eigenspectra
$\espec_j$ and coefficients $\gamma_{ij}$ which correspond to
eigenvalues at or above a threshold $\weiel_m^2$; all the rest
correspond to modes along which the data show relatively small
variation. Our galaxy classification method follows just this
strategy.  

\begin{figure}
\centerline{\epsfxsize=6.0in\epsfbox{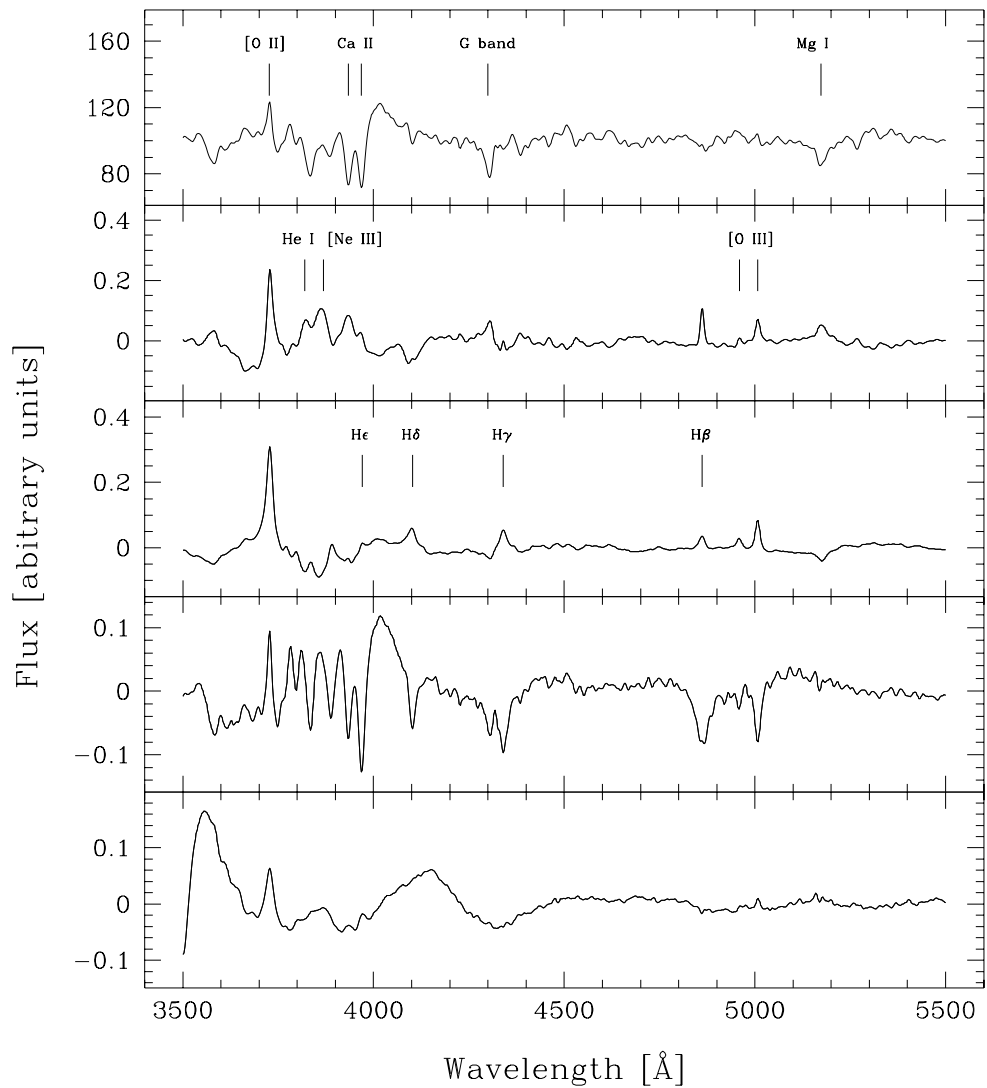}}
\caption{
The mean spectrum of the LCRS galaxies after high-pass
filtering (top), and the first four eigenspectra from singular-value
decomposition.  The first eigenspectrum (which has the largest
eigenvalue) is second from the top, the second eigenspectrum is
immediately below it, and so forth.  The wavelength of some atomic
line features are marked; note that these features represent either
emission or absorption, depending on the sign of the weight
factor multiplying each eigenspectrum.
}
\label{fig:espec}
\end{figure}

\begin{figure}
\centerline{\epsfxsize=6.0in\epsfbox{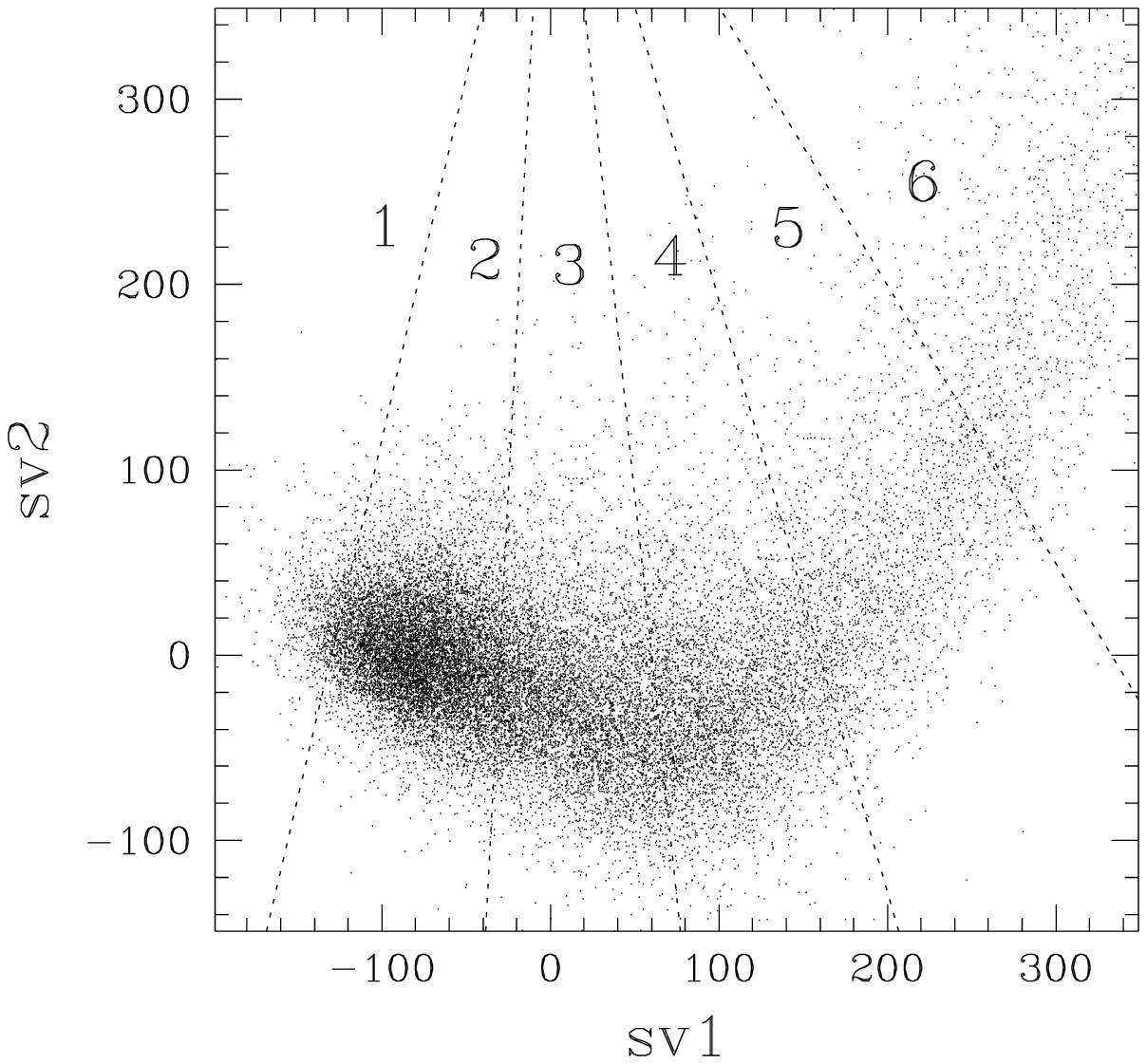}}
\caption{
Scatter diagrams of the projection coefficients (sv1 and sv2) for the first
and second eigenspectra in Figure~1. The regions determined by
the straight lines define the galaxy clans, labeled 1 through 6. 
}
\label{fig:sv1sv2}
\end{figure}

The simplest scheme is to consider only $\gamma_{i1}$, the dot product
of the $i^{\rm th}$ galaxy spectrum with the first eigenspectrum
$\espec_1$. The galaxies have the greatest variance with this
coefficient and it alone may be sufficient to usefully define spectral
type. Then the values of $\gamma_{i1}$ could be partitioned into bins,
with each bin corresponding to a unique type.  Here, we use the first
two eigenspectra (\fig{espec}) to define spectral type because the
geometry of data-point scatter in the vector space of frequency
channels is not exactly elliptical; there exist correlations between
coefficients such that the scatter of data points roughly forms an
arc. We create distinct types by extending boundaries radially outward
from a point at some distance from the arc.  \fig{sv1sv2}\ illustrates
the six spectral types---here called ``clans''---defined in this way.
The boundaries were chosen so that each clan contains roughly similar
pathlengths along a trajectory which runs along the central ridge of
the arching scatter figure.  In this sense the dimensionality of our
classification scheme is one; however, because of the non-Gaussian
nature of the data scatter, the classification parameter's trajectory
lies in a two-dimensional projection of the principal component
eigenvectors.

The choice of clan boundaries as shown in \fig{sv1sv2} were meant to
capture the central peak in the density of the data scatter (clan~2),
and the transition along the ``tail'' of the scatter diagram toward
positive values of projection coefficients (clans 3--5).  The
remaining two clans (1 and 6) are defined to assess the outliers of
the scatter diagram. We find evidence from the systematic trend in
luminosity function shape with clan index (\S\ref{ssect:lfbyclan})
that the outlier clans represent populations of galaxies with unique
properties, as opposed to random scatter of galaxies in neighboring
clans.

Before analyzing astrophysical properties of the clan populations, we
discuss qualitatively the line and continuum features which appear in
the eigenspectra.  This will help place our classification scheme in
the context of morphological classification and other spectral
classification methods.

\subsection{Physical Significance of the Eigenspectra}

Recall that in deriving the eigenspectra, we subtract the mean
spectrum of the 25,327 galaxies. Thus the eigenspectra measure
differential signal in continuum and line features.  \fig{espec}
contains the mean spectrum of the LCRS galaxies along with the first
four eigenspectra.  The first eigenspectrum shows emission features in
H~I, He~I, O~II, O~III, Ne~III, Ca (H~\&~K) and Mg~I, as well as
slight continuum ``absorption'' on the red side of the 4,000\AA\
Balmer limit.  Because the individual spectra are high-pass filtered,
i.e., flattened, the 4,000\AA\ break shows up as a sawtooth-shaped
profile.  Some of the emission features---H$\beta$ and forbidden O~II
and O~III lines---are direct indicators of physical processes,
presumably in the interstellar medium.  Others, such as the Ca~H~\&~K
lines and G-band complex lines, appear in ``emission'' primarily to
compensate for spectra with weaker absorption than is exhibited in the
mean spectrum.

Evidently, the most significant spectral variation between galaxies is
the degree of line emission.  Emission lines of oxygen and hydrogen
Balmer lines are expected from H~II regions, while the diminished
stellar absorption in the calcium lines indicate a young stellar
population. Therefore a positive projection coefficient from this
eigenspectrum seems like a good indicator of star formation.  Indeed,
the most prominent feature in the eigenspectrum, the
[O~II]~$\lambda$3727 line, is known to correlate with star-formation
rate as well as morphological type (Kennicutt~\cite{Ken92a}). However,
there are two caveats. The first is that better measures of star
formation are known, specifically $H\alpha$ flux and the shape of the
UV continuum, but both are outside of the waveband considered
here. The second caveat is that it can be difficult to make a
distinction between star-forming regions and AGNs or LINERs solely on
the basis of the emission lines contained in the eigenspectrum. Again,
H$\alpha$ measurements would have helped (Baldwin, Phillips \&
Terlevich~\cite{BalPhiTer81}; Allen\etal~\cite{AllEtal91}).  The
fraction of objects with non-thermal spectra is expected to be small
and therefore we hold to the interpretation that emission lines
associated with the first eigenspectrum indicate star formation.
Below we present evidence (Fig.~4) that AGN spectra can be isolated,
but using the coefficient of projection with the third eigenspectrum.

The second eigenspectrum is primarily dominated by [O~II] (3727~\AA),
[O~III] (5007~\AA), and H$\gamma$ (4340~\AA) emission lines.  It
allows a fine-tuning of the line contributions in star-forming
galaxies---there is tendency for galaxy spectra which have a large
projection coefficient with the first eigenspectrum to couple with the
second eigenspectrum as well. This correlation is manifested by the
curvature of the data scatter in \fig{sv1sv2}.  The appearance of
strong H$\delta$ emission in the eigenspectrum suggests that it
positively correlates with star-forming H~II regions at relatively
high levels of excitation.

\begin{figure}[p]
\centerline{\epsfxsize=6.0in\epsfbox{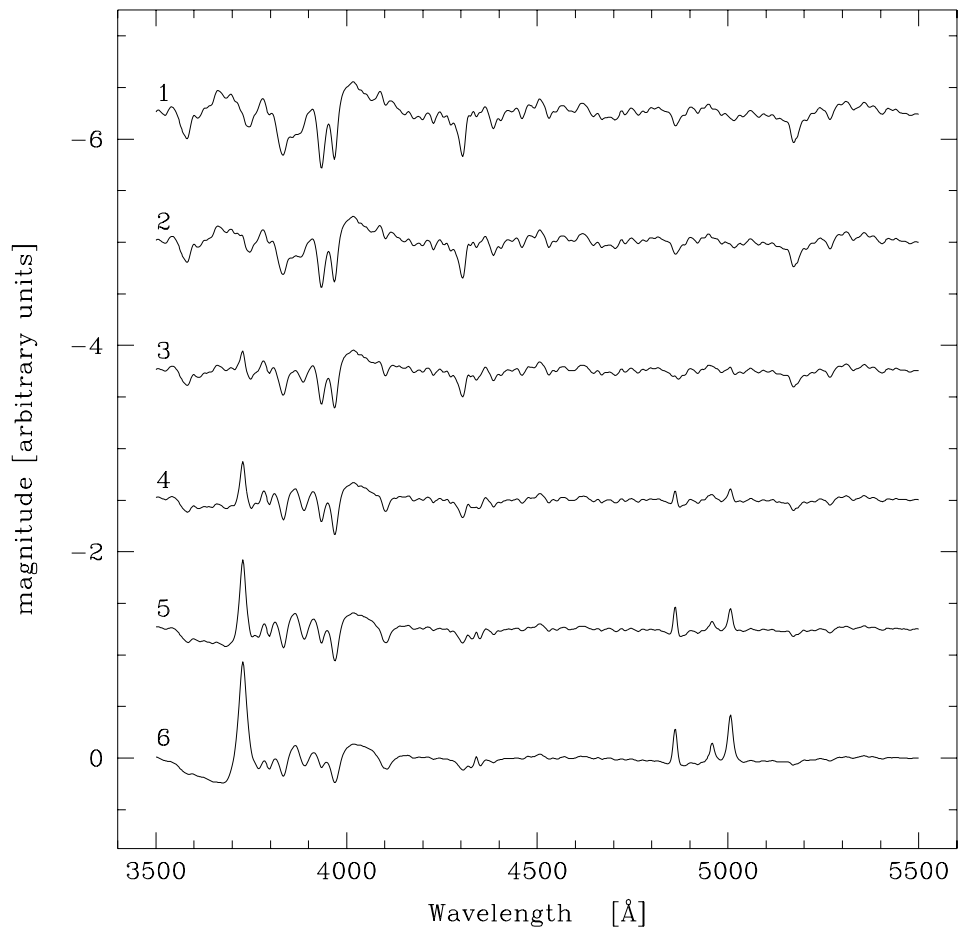}}
\caption{ 
The mean spectra of the six clans. A low-frequency component
has been removed, as in Figure~1, and each curve has been offset for
clarity. The labels indicate the clans as defined in Figure~2.  
\label{fig:clanspec}}
\end{figure}

Ultimately, the physical significance of the eigenspectra is best
interpreted by examining the spectra which they represent.  In
\fig{clanspec} we show the average spectra of objects in the six
clans.  A progression from clan to clan is clear and supports our
interpretation that the most important spectral differences between
galaxies lie in the amount of line emission. 

\begin{figure}[p]
\centerline{\epsfxsize=6.0in\epsfbox{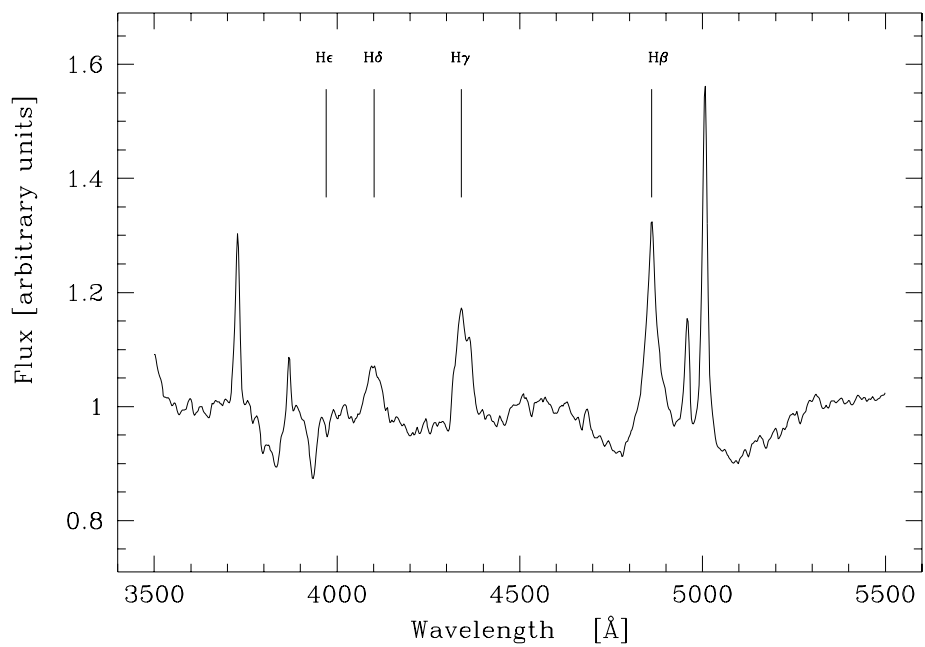}}
\caption{ 
The mean spectrum of galaxies with abnormally strong,
negative projection coefficients for the third eigenspectrum. As in
Figure~3, a low-frequency component has been removed. Strong Balmer
emission is evident in the spectrum, and the interpretation here is
that these are Seyfert~1 galaxies.  
\label{fig:oddones}}
\end{figure}

While not relevant to our present spectral classification scheme, the
third eigenspectrum is noteworthy. It has the role of further
fine-tuning specific line-emission features including
[O~II]~$\lambda$3727, as well as continuum emission above 4000~\AA.
Note that it shows strong absorption in H$\gamma$ and H$\delta$ and an
asymmetry in the Ca~H\&K troughs from H$\epsilon$.  The distribution
of projection coefficients for this eigenspectrum is roughly symmetric
except for a weak tail extending to large negative values, with a
significant population---about 0.6\% of the total---between three and
seven standard deviations. The mean spectrum of these objects is shown
in \fig{oddones}.  The projection coefficients of the third
eigenspectrum are negative for this population, hence it shows strong
emission in the Balmer series. With the pronounced, broad Balmer lines
and enhanced [O~III]/O[~II], we interpret these objects as Seyfert~1
galaxies (cf. Kennicutt~\cite{Ken92b}).

The skewness in the distribution of projection coefficients
steadily decreases with increasing eigenspectrum index. 
By the fourth eigenspectrum, the distribution is nearly
symmetric (skewness is around 0.2, as compared with 1.2 for the
first eigenspectrum). This is reassuring because it suggests
that the scatter in the higher-order eigenspectra is becoming
more and more characteristic of Gaussian scatter, and contains
less information relevant to classification.

\subsection{Comparison with other Spectral Classification Schemes}

A variety of spectral classification schemes for galaxies has been
proposed in the literature, and recent examples include
Allen\etal~(\cite{AllEtal91}), Bershady~(\cite{Ber95}), Zaritsky,
Zabludoff \& Willick (\cite{ZarZabWil95}) and Kurtz \&
Mink~(\cite{KurMin97}). Most of these methods involve fitting to
specific line features or to a predefined spectral template which
includes these features.  Connolly\etal~(\cite{ConEtal95}) proposed a
method which incorporates eigenspectra defined primarily from the data
itself in a strategy very similar to the one used here. We begin this
section by considering their technique.

Connolly\etal\ performed a principal component analysis of optical and
UV data to classify galaxies according to the projection of their
spectra onto eigenspectra.  The eigenspectra were derived from a set
of 10 templates, where each template is the average of spectra from a
given morphological type, or class of starburst galaxy.  Since the
number of frequency channels is large, over 2200 in their case,
Connolly\etal\ properly construct their correlation matrix using the
space defined by the 10 templates, not the space of frequency
channels. Thus each ``data point'' is essentially a measure of how the
flux is distributed among the 10 templates. Connolly\etal\ also assign
a variable weight to each template (as in \eqnp{pcawei}), but
otherwise, the procedure is the same as in a ``standard'' principal
component analysis.  Connolly\etal\ argue that their projection of the
data onto the space of templates offers several advantages over the
standard method, not the least of which is a reduction in
computational load. In general, there is considerable savings in
diagonalizing a $10\times 10$ matrix versus one of size $2200\times
2200$ (although of rank 10).

A nice way to view the differences between the method of
Connolly\etal\ and a standard principal component analysis is through
the formalism of SVD.  The raw data, in this case a $10\times 2200$
matrix formed of 10 vectors in the 2200-dimensional space of frequency
channels, is projected onto the 10-dimensional space of templates
simply with a matrix transpose operator. Indeed, it is formally
necessary to take the transpose of \eqn{svd} to interpret the matrices
$\vee$ and $\wei$ in terms of eigenspectra and eigenvalues as we did
earlier---this enables the data of Connolly\etal\ to meet the
requirement that the number of rows is greater than the number of
columns.  Numerically, it is not at all necessary to work with the
transpose. One simply performs the SVD and pulls the eigenspectra out
from the rows of the matrix $\ewe$ in \eqn{svd} (see
Press\etal~\cite{PreEtal92}, p.~66).

In practice it may not be feasible to decompose a catalog matrix if
the number of data points and the number of frequency channels are
large compared with available machine memory.
%
%
Instead, one may calculate and diagonalize the covariance
matrix directly without ever forming the catalog matrix.
For this purpose we still recommend an SVD algorithm.

Once the eigenspectra were defined, Connolly\etal\ projected
individual galaxy spectra onto them; a classification scheme based on
projection coefficients for the first two eigenspectra produced a
spectral type that was strongly, but not perfectly, correlated with
morphology.  It should be noted that their eigenspectra are most
responsive to the level of broadband continuum flux and to the
discontinuities at 3600~\AA\ and 4000~\AA.  In contrast, our
eigenspectra are not sensitive to broadband continuum flux, as a
result of high-pass filtering. The high-pass filter does admit signal
from the 4000~\AA\ break, which appears as a saw-tooth profile, hence
the continuum break enters our analysis through its variability over a
small number of frequency channels (and not through the absolute
broad-band flux on either side of the break), This suggests that a
benefit of high-pass filtering is sensitivity to key continuum
features with less stringent demands on the accuracy of absolute flux
measurements.

In contrast Zaritsky, Zabludoff \& Willick (\cite{ZarZabWil95})
focussed on line features. They defined a 3-dimensional space based on
ratios of line emission and absorption, projection onto stellar
templates of young (spectral type A) and old (K) stars, and ratio of
hydrogen to oxygen emission lines. A ``training set'' of spectra
enabled the authors to determine the loci of points in this space
which correspond to morphological types. They tested the distance from
these loci as measures of type using an independent set of galaxy spectra, 
and they concur with Connolly\etal\ that the correlation between spectral
and morphological types is strong but not perfect. Like Connolly\etal,
Zaritsky\etal\ make the important point that the lack of perfect
correlation is not the result of flaws in the spectral classification
scheme, but rather reflects an intrinsic property of galaxies.

We emphasize that the present classification scheme is high-pass
filtered, and is more sensitive to line features than the shape of the
continuum.  In this regard our approach is qualitatively similar to
the method of Zaritsky\etal\ Indeed, two of their classification
parameters, the ratio of absorption to emission features and the
relative line strengths of hydrogen and oxygen, correspond directly to
the first two eigenspectra presented in \fig{espec}. The third
parameter used by Zaritsky\etal, a combination of projection 
coefficients with A-star and K-star templates, evidently has
relatively small galaxy-to-galaxy variance (Fig.~3 therein)---this
parameter is not highly effective as a type-discriminator.

Recently, Kurtz \& Mink~(\cite{KurMin97}) devised a classification
scheme which is based on direct estimates of the effects of survey
selection, instrumental response and the astrophysics of line
formation to model template spectra without recourse to a statistical
approach such as the method used here.  They found that their data
could be described simply and accurately by a linear combination of an
absorption spectrum and an emission spectrum, with the following
physical interpretation: All galaxies exhibit the absorption spectrum
generated by their constituent stellar populations, but many also have
superimposed emission features from physically distinct H~II
regions. Thus spectral classes may be derived simply from projection
onto the emission-line template.  A large fraction of the $\sim$2,000
galaxies in their survey have spectra which do not project strongly
onto this template; the peak in the distribution of projection
coefficients for our first eigenspectrum reflects just this behavior,
although in \fig{sv1sv2} a fair amount of scatter is evident about the
peak.

Kurtz \& Mink have thus developed the idea that spectral
classification can be efficiently accomplished with a single parameter
associated with star formation. An early incarnation of this idea came
from Aaronson (\cite{Aar78}) who considered the relative contribution
of red and blue stars in galaxies: To a fair approximation, the mix
of A and K stars can account for variations in the broadband colors of
galaxies (see Bershady \cite{Ber95} for a refinement of this idea).

Despite the existence of essentially one-parameter classification
methods based on the amount of star formation, virtually all schemes,
including the one introduced here, have a working dimensionality of
two or more, reflecting the difficulty in precisely quantifying the
gamut of emission from star-forming regions with a single template. An
alternative might be to project galaxy spectra onto a ``variable
template'' (i.e., a large set of fixed templates) which reflects
subtle changes in spectral profiles as a function of star formation
rate. Here we essentially form such a variable template by working
with a linear combination of the first two eigenspectra.

Given the similarities between our classification scheme and those for
which morphological types have been considered, it is likely that the
clans and morphological type are strongly correlated. Thus we adopt
the nomenclature of ``early'' and ``late'' types to refer to galaxies
solely on the basis of clan membership, with low clan number
corresponding to early types.  We now proceed to consider the
clan-dependence of galaxy properties; the focus here is primarily on
the luminosity function.

\section{Luminosity Functions}\label{sect:lf}

In their comprehensive review, Binggeli, Sandage \& Tammann
(\cite{BinSanTam88}) point out that the luminosity function of
galaxies, $\Phi$, giving the distribution of galaxies as a function of
absolute magnitude $M$, should be defined in terms of galaxy
types. They argue, as have others (Holmberg ~\cite{Hol58} seems to be
the first), that galaxies of different morphological type have
different luminosity functions.  Thus, we write the general luminosity
function of the full LCRS catalog as
\bel{genlf}
\Phi_{g}(M) = \sum_{c = 1}^{6} f_c \Phi_c(M) \ ,
\ee
where $\Phi_c$ is the luminosity function for the $c^{\rm th}$ clan,
and $f_c$ is the fraction of all galaxies of the $c^{\rm th}$ clan in
some specified region of the Universe.  Here we treat all luminosity
functions as probability distributions, hence $\Phi_{g}$ and the
$\Phi_c$ are normalized to unity upon integration over all absolute
magnitude values in the range of interest:
\be
\int^{\Mdim}_{\Mbri} dM \, \Phi(M) = 1 \ ;
\ee
In our analysis $\Mdim = -16.5$ and $\Mbri = -23.0$.

In principle, there can exist a universal luminosity function that
applies to all galaxies without any dependence on local environment. A
case can been made against universality, and we offer some ammunition
for it here.  Binggeli\etal\ provide the starting point, the
assumption that while the general luminosity function is not
universal, the luminosity functions of the individual clans are. Then,
one can argue against universality if the clan luminosity functions
differ from one another and if the relative clan population depends on
local density as is observed (Dressler~\cite{Dre80}).  An
implausible conspiracy would be required to maintain universality when
the luminosity function varies with galaxy and the fraction of each
type varies with local density.  As a first application of the
spectral classification scheme presented here, we address these issues
of type dependence. We begin with an outline of the
methodology.

\subsection{Method of Estimation}

We estimate luminosity functions for the LCRS galaxies using a
procedure which is almost identical to that described in \lcii. First,
we apply the non-parametric estimator proposed by Efstathiou, Ellis \&
Peterson~(\cite{EfsEllPet88}) to obtain the unconstrained form of the
luminosity functions.  Then we fit the data with Schechter functions,
\bel{schechter}
\Phi(M) = C 10^{-0.4 (\alpha+1) (\Mstar-M)} \exp(-10^{0.4 (\Mstar - M)}) 
\ ,
\ee
where $\alpha$ and $\Mstar$ correspond to the slope of the faint-end
of the galaxy distribution and to a typical absolute magnitude,
respectively, and $C$ is a normalization constant,
\be
C = \left. \frac{1}{\Gamma(\alpha+1,10^{-0.4 (\Mstar - M)})}
    \right|^\Mdim_{M  = \Mbri} \ .
\ee
Note that absolute magnitudes are calculated on the basis of
relativistic luminosity distances with cosmological deceleration
parameter $q_0 = 0.5$.  As in \lcii, the results presented here are insensitive 
to the choice of $q_0$ in the range of 0.1 to 0.5. To obtain physical 
quantities such as luminosity and space density, we must also fix the 
Hubble parameter, $H_0 = 100 h$~km/s/Mpc. Where the quantity $h$ does 
not appear explicitly it has been set to unity.

In addition to fitting the luminosity function with a parametrized
form, we assess its ``natural'' shape with a nonparametric method.
The nonparametric representation of choice
is a step-wise function (first-order spline or histogram) of constant
bin width $\delta M$ in magnitude:
\bel{nonparm}
\Phi(M) \approx \sum_{n = -\infty}^{\infty} \phi_n  W_{TH}(M- n \delta\! 
M) \ ,
\ee
where $\phi_n$ are constants, and $W_{TH}$ is a tophat function of
unit height and width, centered at the origin. Of course we cut the
summation so that $n \delta\! M$ lies between $\Mbri$ and $\Mdim$, but
even so, ``nonparametric'' evidently implies a large number of
parameters.

The weights $\phi_i$ in \eqn{nonparm}---or the parameters of any
model of a luminosity function---can be determined from a maximum
likelihood analysis (Sandage, Tammann \& Yahil
~\cite{SanTamYah79}). Define $p_i$ to be the probability of observing
the $i^{\rm th}$ galaxy in a catalog with an absolute magnitude $M_i$,
given that it is located at some specified redshift. This is
equivalent to the fraction of such galaxies that could be observed
given the catalog's minimum and maximum apparent magnitude cuts and
other selection criteria. The likelihood that the full catalog of
galaxies was observed, given their redshifts, absolute magnitudes, the
survey window, is just the
product
\bel{maxl}
{\cal L} = \prod_{i = 1}^N p_i \ .
\ee
The form of the luminosity function can be estimated by first guessing
it and then functionally deforming it to maximize the likehood
function in \eqn{maxl}.  If the number of parameters needed to
determine the luminosity function's form is small, then a direct
search of the parameter space quickly yields the best-fit.  Otherwise,
as in the case of the nonparametric luminosity function in
\eqn{nonparm}, the best fit can be obtained from iterative corrections
using constraints from the derivative of $\cal L$ with respect to the
$\phi_i$.  Uncertainties in the best-fit luminosity function are
calculated either by directly mapping relative confidence levels from
the likelihood function or by calculating elements of the covariance
matrix. Again, see Efstathiou\etal\ or \lcii\ for details.

One virtue of the maximum-likelihood method is that it is insensitive
to the density field---the probability $p_i$ represents the {\em
fraction} of galaxies observable at a given magnitude and this
quantity is independent of density as long as the luminosity function
itself is not density-dependent. We are assuming for the moment that
this is true, at least for the clan luminosity functions.

To complete the estimation of the general luminosity in \eqn{genlf},
we need the number density of galaxies in each clan.  Since density
factors out of the maximum-likelihood estimators, it must be estimated
independently. We use the same prescription as in \lcii\ to get the
mean number density:
\bel{dens}
    n =  \frac{1}{V} \sum_{i = 1}^N \frac{W_i}{\int_{M_+}^{M_-} dM \, 
\Phi(M)} \ ,
\ee
where $V$ is the proper volume of the survey, the integral gives the
luminosity selection function, and $W_i$ is a weight factor to account
for surface-brightness selection effects and variable sampling rates
in the LCRS fields (see \lcii, \S 3.1 therein). The limits of
integration in \eqn{dens}, $M_\pm$, give absolute magnitude cuts which
simultaneously satisfy both the apparent and absolute magnitude limits
of the survey.

\subsection{Luminosity functions by clan}\label{ssect:lfbyclan}

\begin{figure}[p]
\centerline{\epsfxsize=6.0in\epsfbox{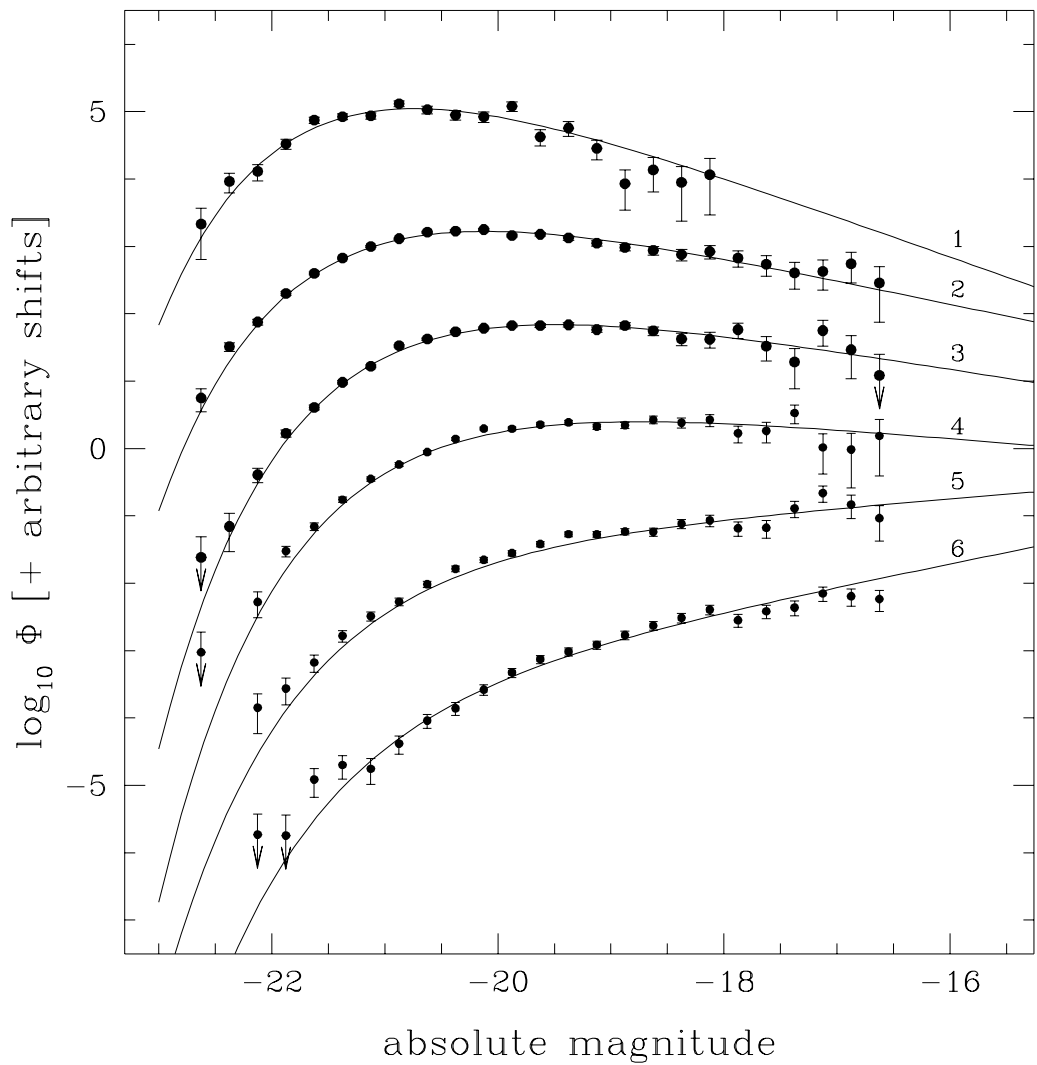}}
\caption{ 
The luminosity function by clan. The points give the
luminosity function from the nonparametric fit and the curves are the
best-fit Schechter functions.  The points and curves have been offset
in the vertical direction for clarity; the labels indicate the clan
identity. Error bars show 1-$\sigma$ confidence limits except in
cases (signified with arrows) where the lower limits are
formally negative. 
\label{fig:sixlf}}
\end{figure}

\begin{figure}[p]
\centerline{\epsfxsize=6.0in\epsfbox{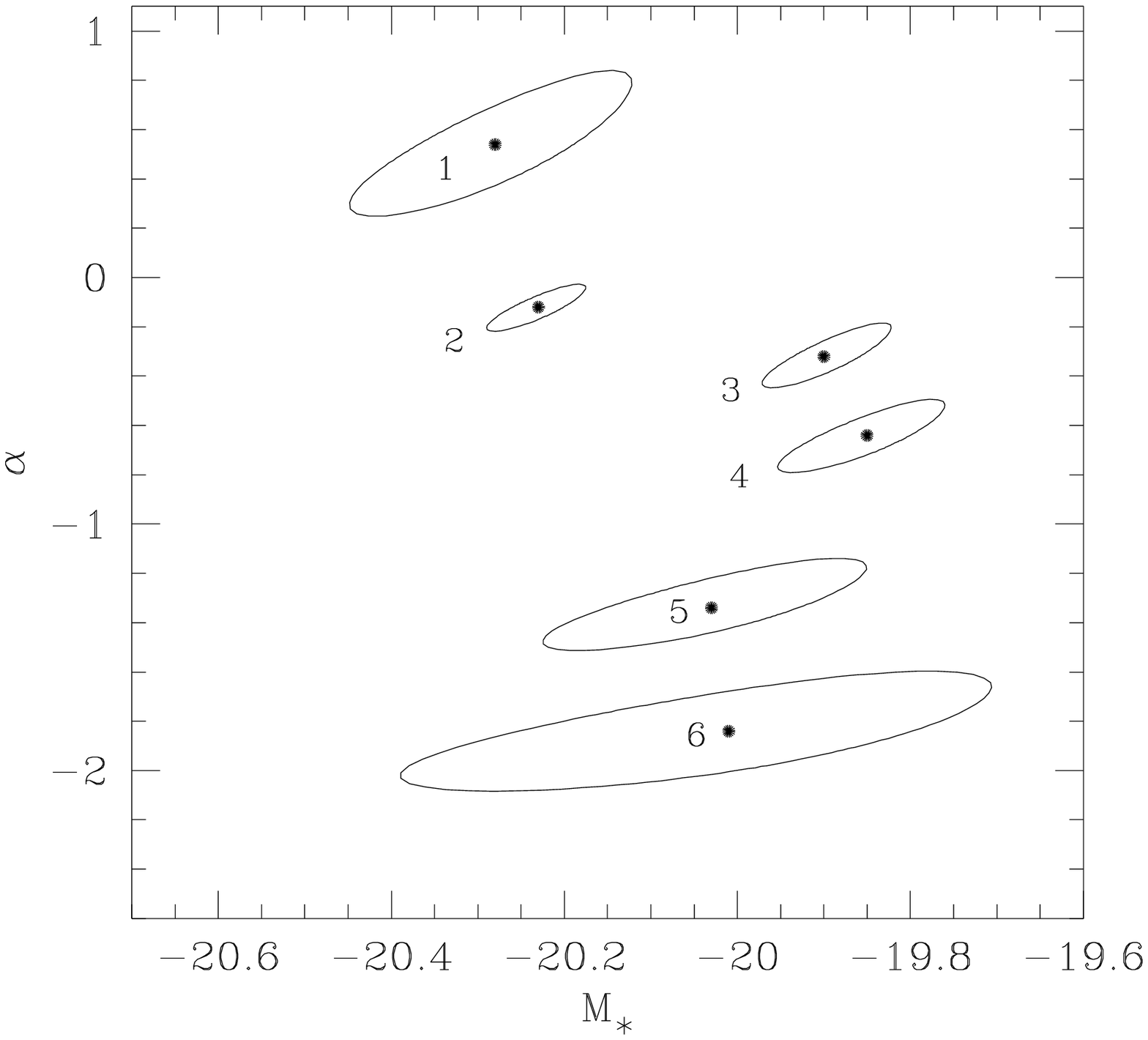}}
\caption{
Error ellipsoids for the best-fit Schechter parameters.  The
labels indicate clan index, the best-fit values of $M_\ast$ and
$\alpha$ are marked with solid circle, and the ellipsoids show 95\
confidence intervals. 
\label{fig:sixlfam}}
\end{figure}

We apply the above procedure to the 6 clans and to the total
population of the LCRS catalog observed with the 112-fiber
spectrometers (NS112); the results are summarized in
\fig{sixlf}, \fig{sixlfam} and Table~\ref{tab:six}.  There is a clear,
systematic steepening of the faint-end slope as the galaxy type
progresses from early to late. The Schechter parameterization is
reasonable over most of the magnitude range considered here, although
the best-fit bright-end falloff may be too steep for late-type
galaxies.  Given that the eigenspectra used in our classification
predominantly measure line emission, the trend in the faint-end slope was
anticipated in \lcii\ from analysis of two subsets of galaxies
partitioned according to the degree of O~II line emission.

\begin{table}[bthp]
\centering
\caption{Luminosity function parameters by clan}
\label{tab:six}
 \centerline{\sc Luminosity function parameters by clan}
\vspace{22pt}
\begin{tabular}{cccccc}
\hline\hline
\rule{0pt}{14pt}Clan & $N$ & $\alpha$ & $\Mstar - 5\log h$ & 
   $\overline{n} \times 10^{-3}$~ h$^3$Mpc$^{-3}$ & 
   $f_{high}^1$ \\
\hline
\rule{0pt}{14pt}
1 & 655  &  $0.54\pm 0.14$ & $-20.28\pm 0.07$ & $0.34\pm 0.03$ & 50.1\
\\ 
2 & 7596 & $-0.12\pm 0.05$ & $-20.23\pm 0.03$ & $7.1\pm 0.6$ & 39.0
\\
3 & 4653 & $-0.32\pm 0.07$ & $-19.90\pm 0.04$ & $9.9\pm 1.3$ & 34.0
\\
4 & 3190 & $-0.64\pm 0.08$ & $-19.85\pm 0.05$ & $11.5\pm 2.1$ & 29.4
\\
5 & 1393 & $-1.33\pm 0.09$ & $-20.03\pm 0.09$ & $8.4 \pm 2.2$ & 29.4
\\
6 & 618 & $-1.84\pm 0.11$ & $-20.01\pm 0.14$ & $13.1 \pm 7.8$ & 29.5
\\
\hline
\end{tabular} 

\vspace{10pt}
{\footnotesize\mbox{}$^1$fraction of objects in each clan which are
   associated with high density regions (see text).
}

\label{tab1}
\end{table}

The formal errors given in Table~1 are derived without making complete
correction for galaxies which were excluded from the LCRS on the basis
of central surface brightness. As discussed in \lci, approximately 9\%
of the galaxies which meet all other selection criteria were cut from
the S112 subsample as result of low central surface brightness. The
remaining galaxies were weighted to compensate for the excluded
objects of similar absolute magnitude. The weights were determined by
comparison with the N112 subsample which has less stringent surface
brightness selection. In the absence of direct measurements, the
results presented here and in \lcii\ are not extrapolated to account
for the 4\% exclusion rate of the N112 subsample. To assess the
systematic errors incurred by following this procedure, we examine the
change in the faint-end slope of the clan luminosity functions from
the N112 subsample with a number of surface-brightness cuts and make
the extrapolation to a cut rate of zero. The results suggest that the
surface brightness cuts only marginally affect the luminosity
functions of the first three clans. For example, the best-fit
faint-end slopes of these functions change by less than $|\Delta
\alpha| = 0.03$. The cuts do affect the luminosity function of the
fainter galaxies in clans 4, 5 and 6 to a greater degree; the
faint-end slopes are expected to shift systematically downward by
0.07, 0.12 and 0.05, respectively.

\begin{figure}[p]
\centerline{\epsfxsize=6.0in\epsfbox{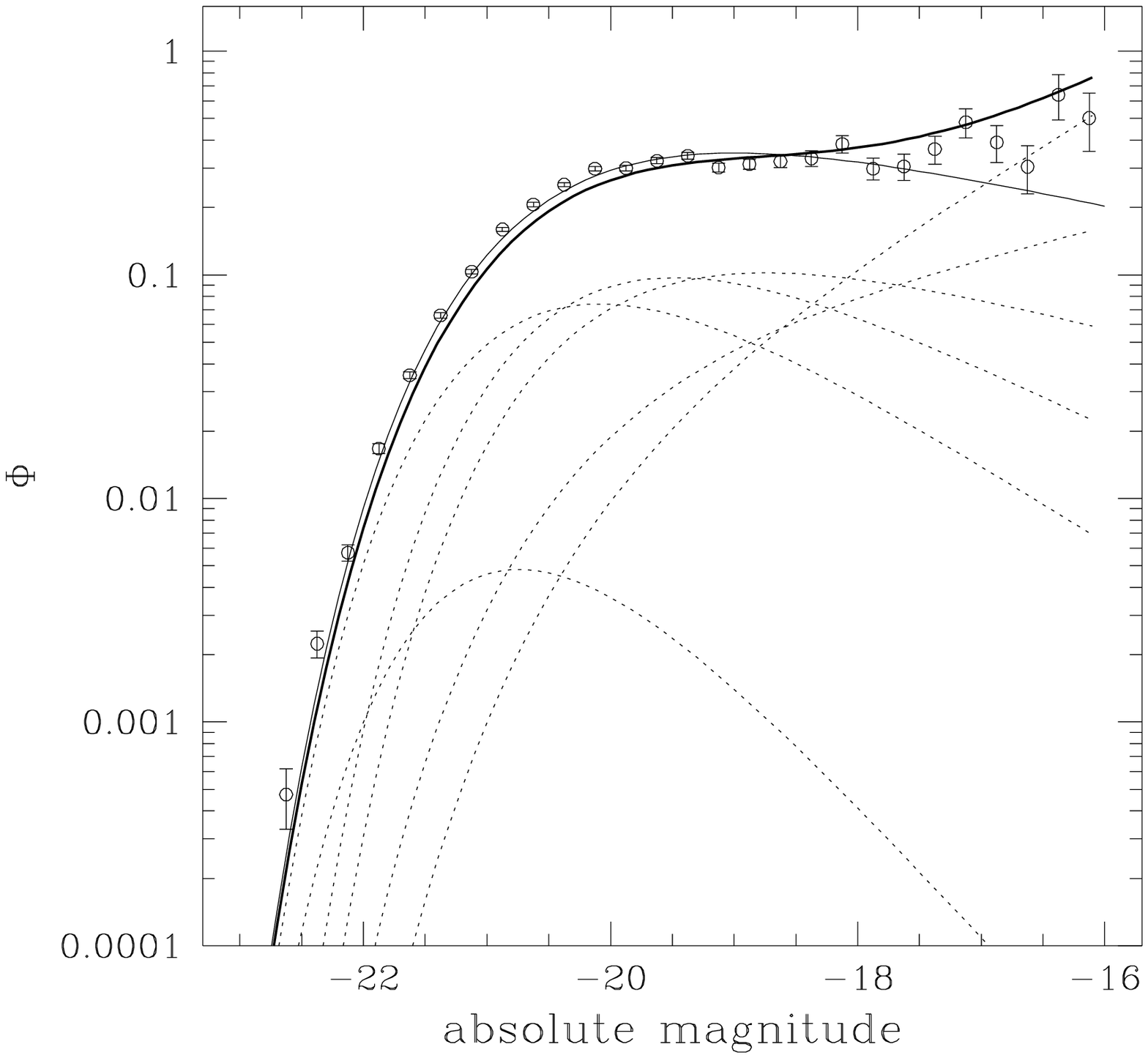}}
\caption{ 
The general luminosity function. The points give the
luminosity function from the nonparametric fit for the full, NS112
catalog, and the light solid curve is a Schechter function fit to the
data.  The heavy solid curve is the general luminosity function in
equation~(6); the individual clan luminosity functions which
contribute to it are shown as well (light dotted lines), weighted
according to their contribution.  The clan identity of the dotted
curves can be easily determined from the faint-end slope (cf. Figure
5).  
\label{fig:genlf}}
\end{figure}

The 
effectiveness of the Schechter parameterization tempts us to fit not
only the clan luminosity functions but the universal luminosity
function for the whole catalog as well.  \fig{genlf} compares the
best-fit universal luminosity function from \lcii\ with a weighted sum
of fitted clan luminosity functions, as in \eqn{genlf}.  Note that the
faint-end slope of the \lcii\ fit is relatively flat, with $\alpha
\sim -0.7$.  In contrast, the faint-end slope of the weighted sum is
formally $\alpha \sim -1.8$, which corresponds to the very late-type
(clan 6) galaxies. If one were to pull the faint-end slope from the
behavior of the weighted sum in the range --18.5 to --17.5~Mag, then
the slope would be approximately $-1.1$, a value which is more in line
with other redshift surveys (e.g., Marzke\etal~\cite{MarEtal94}), as
discussed below in \S\ref{ssect:compRSs}. We conclude that the
Schechter parameterization of the general luminosity function is
robust over a limited magnitude range near $\Mstar$ but can give
highly misleading results when extrapolated to faint magnitudes.

A refinement of the analysis that considers northern and southern
populations separately preserves the general trend of faint-end slope
with galaxy type.  However, some of the clans, in particular clans 1
and 3, show significant differences in the northern and southern
subsamples.  A similar situation is discussed in \lcii\ for the
estimation of a universal luminosity function from the full NS112
data: the error ellipsoids in the $\alpha$--$\Mstar$ plane overlap
only at the 2-$\sigma$ level.  It is argued in \lcii\ that the problem
arises not from the faint galaxies but from the bright ones, which can
modify $\alpha$ if the Schechter parameterization gives only an
approximate fit.  This would suggest that the surface brightness
selection of the 112-fiber data is not the cause of the discrepancy.
Another possible source of the discrepancy is cosmic variance from
large-scale structure and environmental effects. We examine this topic
next, but we must bear the north-south differences in mind.

\subsection{Environmental dependence}\label{ssect:densdep}

The density-morphology relation (Dressler~\cite{Dre80}; Postman \&
Geller \cite{PosGel84}) quantifies the extent to which the mix of
galaxy types changes as a function of local density.  To examine this
effect in the LCRS, its galaxies may be divided into high- and
low-density subsets using the 3-dimensional redshift-space
information. Of the many possible ways of making this division we opt
for a friends-of-friends algorithm (Huchra \& Geller \cite{HucGel82})
to identify groups of objects in redshift-space.  The objective here
is to simply but unambiguously determine high density regions.  We do
not weigh the advantages and disadvantages of this choice except to
say that while the method may misclassify some galaxies in small,
high-density groups, it handles redshift-space
distortions---particularly ``fingers of God''---very well. With link
parameters of 75~km/s in the plane of the sky and 500~km/s along the
line of sight and a 4-object minimum group size, the algorithm
identifies about 35\% of the total survey as grouped galaxies. Note
that the 50-fiber objects are included in the group-finding step so
that we can link members of superclusters which may span more than one
$1.5^\circ \times 1.5^\circ$ 112-fiber field in the survey.

Once tagged, the objects in high-density regions are tallied and
compared with their low-density counterparts. The results from raw
number counts show clearly the density-morphology relation; the last
column of Table~1 contains the fraction of 112-fiber members within
each clan that inhabit high-density environments. Over half of the
extreme early-type galaxies (clan 1) are found in regions of high
density, whereas these regions contain less than a quarter of the
extreme late-type (clan 6) objects. The changing mix of galaxy types
with local density, although only crudely established here, clearly
suggests that the general luminosity function is not universal.

The analysis may be taken a step further to determine if the
luminosity functions for the individual types are universal. Using the
same partition as above, we find that there is a significant
density-dependence in the faint-end slope, at least for the early-type
objects. The effect is quite pronounced in the clan-2 galaxies, with
$\alpha = 0.19 \pm 0.07$ in low-density regions and $-0.40 \pm 0.07$ in
high-density regions. This effect and the density-morphology relation
are considered in greater depth in a separate paper.

Incidentally, we have recalculated the general luminosity function
$\Phi_{g}(M)$ taking into account the variation of the type-specific
luminosity functions in high- and low-density regions.  The result
supports the conjecture by Binggeli, Sandage \& Tammann
(\cite{BinSanTam88}) that the assumption of universality for the
individual type-specific luminosity functions is still a reasonable
approximation for estimating $\Phi_{g}(M)$.

\subsection{Comparison with other catalogs}\label{ssect:compRSs}

In many analyses of the luminosity function, the faint-end slope
$\alpha$ from the Schechter parameterization has provided the means to
compare different galaxy populations.  In several recent examples, the
value of $\alpha$ has been reported to be about --1.  For example,
Loveday\etal\ (\cite{LovEtal95}) find $\alpha = -0.97 \pm 0.15$ from
the optically selected Stromlo-APM Redshift Survey,
while Marzke, Huchra \& Geller~(\cite{MarGelHuc94}) report $\alpha =
-1.0 \pm 2.0$ in the CfA Redshift Survey.  The Autofib Redshift Survey
of 1700 blue-selected galaxies (Ellis\etal\ \cite{EllEtal96}) and the
red-selected Century Survey (Geller\etal\ \cite{GelEtal97}) both give
similar values, with $\alpha \approx -1.1$ and $\alpha \approx -1.2$
respectively.

As in the case of the LCRS, there is evidence that these reported
values do not reflect the true behavior of the faint end of the
general luminosity function. Marzke\etal~(\cite{MarEtal94}) identify a
population of Sm and Im galaxies in the CfA Survey which is expected
to dominate the faint end with a slope of $\alpha \sim -1.9 \pm 0.2$.
More recently, Sprayberry\etal\ (\cite{SprEtal97}), in their
examination of blue-selected low-surface brightness galaxies, find a
best-fit Schechter parameter of $\alpha \approx -1.4$.  They also
employ a model consisting of a Schechter function for magnitudes
brighter than a fixed cut-off value and a power law at fainter
magnitudes; the power law is constrained to match the Schechter
function at the cut-off point. The power-law index $\beta$ was
determined to be $\beta = -2.20$, although taken alone, the three data
points which constitute ``the faint end'' do not evidently track such
a steep power law (see Fig.~6 therein).  Nonetheless, the data do
suggest a significant excess of faint, low-surface-brightness galaxies
over the predictions from a fit to the Schechter parameterization.
Similarly Zucca\etal\ (\cite{ZucEtal97}) demonstrate that the faint
end of the ESO Slice Project redshift survey is better represented by
a pure power law of index $\beta \approx -1.6$ which differs
significantly from the value of $\alpha \approx -1.2$ that is inferred
from fitting the luminosity function near $\Mstar$.  They find that
this slope steepens to $\beta \approx -1.7$ for galaxies with emission
lines.

Additional evidence for a steep faint-end slope comes from
Driver\etal\ (\cite{DriEtal95}) who measure galaxy number counts as a
function of apparent magnitude in a deep image from the {\em Hubble
Space Telescope} wide-field planetary camera.  They model the number
count distribution of three groupings of galaxy types using the
Schechter parameterization and argue that late-type objects require
some combination of a steep ($\alpha \lae$~--1.5) faint-end slope and
evolution of the luminosity function in an attempt to reconcile both
the (high) number counts and the (low) local density of faint objects
(e.g., Phillips \& Driver \cite{PhiDri95}).

There is no strong signal of a faint-end up-turn in the general
luminosity function of the Century Survey as determined by
Geller\etal\ (\cite{GelEtal97}), just as in the type-independent
analysis of the LCRS in \lcii. This is reassuring given the
similarities between the waveband selection of the two surveys. Yet
despite these similarities, there are significant differences in the
best-fit Schechter parameters. The problem is likely triggered in part
by the use of the Schechter parameterization in the first place: The
results of \S\ref{ssect:lfbyclan} suggest that the parameterization is
not well-suited to describing the general luminosity function except
over a narrow range of magnitudes. For example, relatively small
changes in the absolute magnitude range used in a fit can
significantly alter the inferred parameters.  A simple change in the
{\em bright} magnitude limit from $\Mbri = -23$ to $-21$ {\em
decreases} the best-fit $\alpha$ for the LCRS from --0.7 to --0.85.
This example suggests that the discrepancy between the LCRS and the
Century Survey seems to lie at the bright end of the luminosity
function, and is not the result of missing faint galaxies.

Another explanation of the discrepancy between the LCRS and Century
Survey results might come from the fact that the Century Survey
encompasses the Corona Borealis supercluster and therefore may have a
different mix of galaxy types as a result of the density-morphology
relation.  A better comparison with the LCRS results will be enabled
with the forthcoming type-dependent analysis of the Century Survey
(Kurtz\etal~1998).

\section{Conclusion}

In this paper we have implemented a spectral classification scheme
based on singular-value decomposition. The SVD approach isolates the
most significant variations in the spectra of a catalog with a set of
orthonormal basis vectors, or eigenspectra. Projection of individual
galaxy spectra onto the two most significant eigenspectra yields
coefficients from which galaxy types are determined.  The method
was applied to the Las Campanas Redshift Survey, and in that catalog
we defined 6 spectral classes (here called clans). The physical
properties of each clan were interpreted in terms of star formation;
the known correlation between star-forming galaxies and morphological
type (Kennicutt 1992) suggests that the clan index runs smoothly from
early to late types.

The eigenspectra derived from singular-value decomposition are also
useful for finding galaxies with unusual spectral properties. The
projection coefficients of the third eigenspectrum alone allowed us to
isolate a distinct population of objects which we identify as
Seyfert~1 galaxies (0.6\% of the total).

As an application of the classification scheme presented here, we used
spectral types to study the luminosity function of galaxies. A strong
type-dependence is identified, as manifested by the faint-end slope,
with $\alpha$ ranging from about $+$0.5 (early types) to --1.8 (late
types).  We include this type-dependence in the calculation of the
general luminosity function and we identify a faint-end slope which
rises steeply, with a power-law index of $\sim -1.8$ in the
faint limit.  Furthermore, the general luminosity function exhibits a
broad shoulder at an absolute magnitude of roughly --20 in the
$R$-band. These two features are not both represented by the
Schechter parameterization.  On the other hand, the individual
type-specific luminosity functions are mostly well-characterized by the
Schechter parameterization. Small deviations from the
model occur for the luminosity function of the extreme late-type
galaxies, which show a slight excess at bright magnitudes.

The LCRS galaxies are also seen to reflect the density--morphology
relation (cf. Table~1, last column). Given the wide range of
luminosity function shapes among the six galaxy types considered here, any
type-independent assessment of the luminosity function may be
sensitive to the large-scale structure encompassed by the survey. In
addition we report preliminary evidence for variation in the
type-specific luminosity function of some clans as a function of local
density. The effect is most prominent in the (early-type) clan 2
population.

Finally we compared the LCRS luminosity function with those of other
redshift surveys. The case for a steep faint-end rise to the
luminosity function, with a power-law index around $-2$, is now quite
strong. The lower values of faint-end slope reported in the literature
result primarily from poor sampling of faint galaxies, coupled with an
inappropriate extrapolation of the best-fit $\alpha$ parameter. In
some recent work (e.g., Zucca\etal\ \cite{ZucEtal97}, Sprayberry\etal\
\cite{SprEtal97}) the excess number of faint galaxies has prompted the
use of a model consisting of a Schechter function with a break at some
faint magnitude to a pure power law.  A similar effect can be achieved
by taking the luminosity function to be a linear combination of
Schechter functions with varying $\alpha$ parameters, and on the basis
of the results presented here, we argue that this latter approach is
more realistic.

The success of the spectral classification method used here on the
LCRS galaxies suggests that it may be fruitfully applied to larger
catalogs such as the upcoming Sloan Digital Sky Survey. Then, with
better sky coverage we may be able to examine in greater detail the
luminosity function, a critical ingredient for understanding
cosmic structure formation.

\acknowledgements

We thank M.~Kurtz for helpful discussions regarding spectral
classification, and the referee, M.~Bershady, for many insightful
comments and suggestions which greatly enhanced the presentation and
focus of this paper.  WHP and BCB acknowledge funding from NSF Grant
PHY 95-07695.  BCB is grateful to NASA Offices of Space Sciences,
Aeronautics, and Mission to Planet Earth for providing computing
resources.

\end{document}